\documentclass[nofootinbib,twocolumn,preprintnumbers]{revtex4-1}
\usepackage{amsmath} 
\usepackage{pstricks}
\usepackage{longtable}
\usepackage[dvips]{graphicx}       
\usepackage{color}
\usepackage{amssymb}

\input{epsf}

\newcommand\as{\alpha_{\mathrm{S}}} 
\newcommand\f[2]{\frac{#1}{#2}}

\def\beq{\begin{equation}} 
\def\eeq{\end{equation}} 
\def\to{\rightarrow} 
\def\nn{\nonumber}

\def\b0{\beta_0}

\def\beeq{\begin{eqnarray}}
\def\eeeq{\end{eqnarray}}
\def\ep{\epsilon}

\def\mur{\mu_R} 
\def\muf{\mu_F}
\def\mur2{\mu_R^2} 
\def\muf2{\mu_F^2}

\begin{document}

\begin{titlepage}
\renewcommand{\thefootnote}{\fnsymbol{footnote}}
\begin{flushright}
DOI: 10.1103/PhysRevLett.111.201801
\end{flushright}

\begin{center}
{\Large \bf

Higgs Boson Pair Production at\\
Next-to-Next-to-Leading Order in QCD
}
\end{center}
\par \vspace{2mm}
\begin{center}
{\bf Daniel de Florian}\footnote{deflo@df.uba.ar} and
{\bf Javier Mazzitelli}\footnote{jmazzi@df.uba.ar}\\

\vspace{5mm}

Departamento de F\'\i sica, FCEyN, Universidad de Buenos Aires, \\
(1428) Pabell\'on 1, Ciudad Universitaria, Capital Federal, Argentina\\

\end{center}

\begin{center} {\large \bf Abstract} \end{center}
\begin{quote}
\pretolerance 10000

We compute the next-to-next-to-leading order QCD corrections for standard model Higgs boson pair production inclusive cross section at hadron colliders within the large top-mass approximation.
We provide numerical results for the LHC, finding that the corrections are large, resulting in an increase of ${\cal O}(20\%)$ with respect to the next-to-leading order result at c.m. energy $\sqrt{s_H}=14\,\text{TeV}$.
We observe a substantial reduction in the scale dependence, with overlap between the current and previous order prediction.
All our results are normalized using the full top- and bottom-mass dependence at leading order.
We also provide analytical expressions for the $K$ factors as a function of $s_H$.

\end{quote}

\end{titlepage}

\setcounter{footnote}{1}
\renewcommand{\thefootnote}{\fnsymbol{footnote}}

\section{Introduction}

The recent discovery of a new boson \cite{Aad:2012tfa,Chatrchyan:2012ufa}, so far compatible with the long sought standard model (SM) Higgs boson \cite{Englert:1964et}, at the Large Hadron Collider (LHC) opens a new stage in the task of understanding the mechanism of electroweak symmetry breaking.
In order to determine the connection between this phenomenon and the new particle, it is crucial to measure its couplings to gauge bosons, fermions and its self-interactions.
In particular, the knowledge of the Higgs self-couplings is the only way to reconstruct the scalar potential.

Higgs trilinear coupling can be studied via Higgs pair production. 
Recently, several papers have analyzed the possibility of measuring this process at the LHC \cite{Baur:2002qd,
Dolan:2012rv,Papaefstathiou:2012qe,
Baglio:2012np,Baur:2003gp,Dolan:2012ac,Goertz:2013kp,
Shao:2013bz,Gouzevitch:2013qca}.
In general, it has been shown that despite the smallness of the signal and the large background its measurement can be achieved at a luminosity upgraded LHC.
For example, for $b\bar{b}\gamma\gamma$ and $b\bar{b}\tau^+\tau^-$ final states, after the application of proper cuts, the significances obtained are $\sim 16$ and $\sim 9$ respectively, for a c.m. energy of $14\,\text{TeV}$ and an integrated luminosity of $3000\,\text{fb}^{-1}$ \cite{Baglio:2012np}. These are, so far, the most promising final states for the Higgs trilinear coupling analysis.
The sensitivity of these channels can be further improved by the application of jet substructure techniques, as it was shown in Refs. \cite{Dolan:2012rv,Papaefstathiou:2012qe,Gouzevitch:2013qca}.

The SM Higgs pair production at hadron colliders is dominated by the gluon fusion mechanism mediated by a heavy-quark loop.
At leading order (LO) in QCD perturbation theory this process can occur either through a box $gg\to HH$ or a triangle $gg\to H^*\to HH$ diagram, of which only the latter is sensitive to the Higgs trilinear coupling. This cross section has been calculated in Refs. \cite{Glover:1987nx,Eboli:1987dy,Plehn:1996wb}.
The QCD next-to-leading order (NLO) corrections, within the large top-mass ($M_t$) approximation, have been computed in Ref. \cite{Dawson:1998py}, finding an inclusive {\it K} factor close to $2$.
The size of this correction makes essential to reach higher orders to be able to provide accurate theoretical predictions.

Recently, the two-loop corrections were calculated by us in Ref. \cite{deFlorian:2013uza}, again in the large top-mass  limit, and the next-to-next-to-leading order (NNLO) cross section was evaluated within the soft-virtual approximation, following the results of Ref. \cite{deFlorian:2012za}. We found an increase close to $23\%$ with respect to the NLO result.

The finite top-mass effects were analyzed at NLO in Ref. \cite{Grigo:2013rya}, finding that the accuracy of the large top-mass approximation at NLO is dramatically improved if the exact top-mass leading order cross section is used to normalize the corrections, achieving a precision of ${\cal O}(10\%)$.

In this article we present the full NNLO corrections for the inclusive cross section in the large top-mass limit. We also provide numerical predictions for the LHC, using the exact leading order result to normalize the partonic cross section.

\section{Results}

Within the large top-mass approximation, the effective single and double-Higgs coupling to gluons is given by the following Lagrangian
\beq
{\cal L}_{\text{eff}}=
-\f{1}{4}G_{\mu\nu}G^{\mu\nu}
\left(C_H\f{H}{v}-C_{HH}\f{H^2}{v^2}\right)\,,
\eeq
where $G_{\mu\nu}$ stands for the gluonic field strength tensor and $v\simeq 246\,\text{GeV}$ is the Higgs vacuum expectation value.
While the ${\cal O}(\as^3)$ of the $C_{H}$ expansion is known \cite{Kramer:1996iq,Chetyrkin:1997iv}, the QCD corrections of $C_{HH}$ are only known up to ${\cal O}(\as^2)$ \cite{Djouadi:1991tk}. Up to that order, both expansions yield the same result.
Even when this approximation is not reliable at LO, it is a very accurate mechanism for the computation of the higher order corrections if the exact LO result is used, since QCD corrections are dominated by soft contributions which are not affected by the details of the effective vertex.

To compute the SM Higgs boson pair production cross section to NNLO accuracy, we need to evaluate the QCD perturbative expansion up to ${\cal O}(\as^4)$.
We will separate the contributions to the squared matrix element into two classes: (a) those containing two gluon-gluon-Higgs vertices (either $ggH$ or $ggHH$) and (b) those containing three or four effective vertices. Then the partonic cross section will be written as
\beq
Q^2\f{d\hat\sigma}{dQ^2}=\hat\sigma^a + \hat\sigma^b\,,
\eeq
where $Q^2$ is the squared invariant mass of the Higgs pair system.
For the sake of completeness we also include the LO and NLO contributions in $\hat\sigma^a$ and $\hat\sigma^b$.

Contributions to $\hat\sigma^a$ only contain diagrams with one effective vertex each. Given the similarity between $ggH$ and $ggHH$ vertices, the corrections are equal to those of single Higgs production \cite{Harlander:2002wh,Anastasiou:2002yz,Ravindran:2003um} up to an overall LO normalization. Specifically, for each partonic subprocess $ij\to HH+X$ we have (for factorization and renormalization scales $\mu_F=\mu_R=Q$)
\beeq
\label{sigma_a}
\hat\sigma^a_{ij} &=& \hat\sigma_{\text{LO}}\,
\bigg\{
\eta_{ij}^{(0)}
+\left(\f{\as}{2\pi}\right)2\,\eta_{ij}^{(1)}
+\left(\f{\as}{2\pi}\right)^2\bigg[4\, \eta_{ij}^{(2)}\\
&&+\,8\,\delta_{ig}\delta_{jg}\delta(1-x) \f{\text{Re}(C_{LO})}{\left\vert C_{LO} \right\vert ^2}(C_{H}^{(2)}-C_{HH}^{(2)})
\bigg]
\bigg\}\,,\nn
\eeeq
where
\beq
\hat\sigma_{\text{LO}}=\int_{t_-}^{t_+}\!\!dt
\f{G_F^2\, \as^2}{512(2\pi)^3}
\left\{
\left| C_\triangle F_\triangle + C_\square F_\square \right|^2
+ \left| C_\square G_\square \right|^2
\right\}\,,
\eeq
and for the sake of brevity we refer the reader to Ref. \cite{Anastasiou:2002yz} for the expressions of $\eta_{ij}$ and to Ref. \cite{Plehn:1996wb} for $C_\triangle$, $F_\triangle$, $C_\square$, $F_\square$ and $G_\square$.
The term proportional to $\delta_{ig}\delta_{jg}$ in Eq. (\ref{sigma_a}) arises from a possible difference between the second order corrections to the vertices $ggH$ and $ggHH$, of which the latter is still unknown ($C_{HH}^{(2)}$ and $C_{H}^{(2)}$ are defined as in Ref. \cite{deFlorian:2013uza}).
The exact LO partonic cross section $\hat\sigma_{\text{LO}}$ depends on $Q^2$ and $t$,  the latter given by
\beq
t=-\f{1}{2}\left(Q^2-2M_H^2- Q\sqrt{Q^2-4M_H^2}\cos\theta_1\right)\,,
\eeq
where $\theta_1$ is the scattering angle in the Higgs pair c.m. system and the integration limits $t_\pm$ correspond to $\cos\theta_1=\pm 1$.
In the large top mass limit $\hat\sigma_{\text{LO}}$ takes the following simple form,
\beq
\hat\sigma_{\text{LO}}=
\int_{t_-}^{t_+}\!dt
\left(\f{\as}{2\pi}\right)^2
F_{LO} \left| C_{LO} \right|^2\,,
\eeq
where
\beq
F_{LO} = \f{G_F^2}{2304\pi}\,,
\;\;\;\;\;
C_{LO} = \f{3 M_H^2}{Q^2-M_H^2+iM_H\Gamma_H}-1\,.
\eeq
Here $\Gamma_H$ stands for the Higgs total width, while $G_F$ is the Fermi coupling.

Since each $ggH$ and $ggHH$ vertex is proportional to $\as$, contributions to $\hat\sigma^b$ first appear at NLO, as a tree-level contribution to the subprocess $gg\to HH$. Then at NNLO we have one-loop and single real emission corrections. The former have been calculated in Ref. \cite{deFlorian:2013uza}. Specifically, they are all the terms of Eq. (8) of Ref. \cite{deFlorian:2013uza} which are not proportional to $\vert C_{LO} \vert ^2$, except for the term proportional to $(C_{H}^{(2)}-C_{HH}^{(2)})$ which we have already moved into $\hat\sigma^a$.
We will denote this contribution by $\hat\sigma^{(v)}$.

\begin{figure}
\begin{center}
\begin{tabular}{c}
\epsfxsize=8truecm
\epsffile{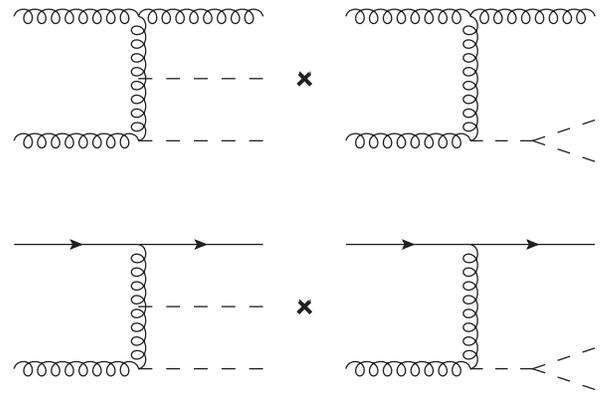}\\
\end{tabular}
\end{center}
\vspace{-0.5cm}
\caption{\label{diagramas}
Example of Feynman diagrams needed for the NNLO calculation for $gg\to HHg$ (top) and $qg\to HHq$ (bottom) subprocesses.  Other parton subprocesses can be obtained from crossings.}
\end{figure}

Finally, the only remaining part of the NNLO contribution to the cross section arises from the real emission processes present in $\hat\sigma^b$, which we will denote by $\hat\sigma^{(r)}$. The partonic subprocesses involved are $gg\to HH+g$ and $qg\to HH+q$ (with the corresponding crossings). Examples of the Feynman diagrams involved in the calculation are shown in Figure \ref{diagramas}.

To compute this contribution we used the Mathematica packages FeynArts \cite{Hahn:2000kx} and FeynCalc \cite{Mertig:1990an} in order to generate the Feynman diagrams and evaluate the corresponding amplitudes.
The calculation was performed using nonphysical polarizations, which we cancelled including ghosts in the initial and final states.
The results for $n=4$ dimensions can be found in the appendix.
In order to subtract the soft and collinear divergencies, we used the Frixione, Kunszt, and Signer subtraction method \cite{Frixione:1995ms}. Below, we provide the details of the calculation.

Let $p_1$ and $p_2$ be the momenta of the incoming partons, $k_1$ and $k_2$ the momenta of the Higgs bosons and $k$ the momentum of the outgoing parton.
We define the variables $x$ and $y$, where $x=Q^2/s$ and $y$ is the cosine of the angle between $p_1$ and $k$. In terms of these variables soft singularities correspond to the limit $x\to 1$, while collinear singularities arise when $y\to \pm 1$.
Gluon initiated subprocesses contain the three kinds of singularities while those processes initiated by quark-gluon are only affected by a collinear singularity and quark-antiquark ones are finite.

The three-particle phase space (PS) in $n=4-2\ep$ dimensions is given by
\beeq
d\text{PS}_3&=&
(4\pi)^{-2+\ep}\f{\Gamma(1-\ep)}{\Gamma(1-2\ep)}\,
d\text{PS}_2^{(x)}\, \f{s^{1-\ep}}{2\pi}\\
&\times&(1-x)^{1-2\ep} (1-y^2)^{-\ep}
dy\, \sin^{-2\ep}\theta_2\, d\theta_2\,,\nn
\eeeq
where $d\text{PS}_2^{(x)}$ is obtained from the two-particle phase space through the replacement $s\to x\,s$, that is
\beeq
d\text{PS}_2^{(x)}&=&
\f{(16\pi)^{-1+\ep}}{\Gamma(1-\ep)}\, (x\,s)^{-\ep}
\left(1-\f{4M_H^2}{x\,s}\right)^{\tfrac{1}{2}-\ep}\\ 
&\times&\sin^{-2\ep}\theta_1\, d\cos\theta_1 \,dx\,,\nn
\eeeq
The variables $\theta_1$ and $\theta_2$ are the polar and azimuthal angles of the Higgs boson with momentum $k_1$ in the center of mass frame of the Higgs pair system, and both of them range between $0$ and $\pi$.
All the relevant invariants for the process can be expressed in terms of $x$, $y$, $\theta_1$ and $\theta_2$ (see the appendix). For more details about this parametrization see, for example, Ref. \cite{Frixione:1993yp}.

We will focus now on the $gg\to HH+g$ subprocess since it suffers from all kinds of singularities. In the soft limit, the squared matrix element has a divergent behavior proportional to $(1-x)^{-2}$, while in the collinear limits it goes like $(1-y^2)^{-1}$.
Combining those factors with the ones coming from the phase space, there is an overall factor $(1-x)^{-1-2\ep}(1-y^2)^{-1-\ep}$ which regularizes all the divergences.
The key to isolating the singularities, then, is to perform the $\ep$ expansion of that factor in the following way \cite{Frixione:1993yp}:
\beeq\label{expansion}
&&(1-x)^{-1-2\ep}(1-y^2)^{-1-\ep}=
-\f{1}{2\ep}\,\delta(1-x)\,(1-y^2)^{-1-\ep}\nn\\
&&\;\;\;\;\;\;-\f{2^{-2\ep}}{2\ep}\left[
\delta(1-y)+\delta(1+y)\right]\\
&&\;\;\;\;\;\;\;\;\;\;\times\left[\left(\f{1}{1-x}\right)_+ -2\ep\left(\f{\log(1-x)}{1-x}\right)_+\right]\nn\\
&&\;\;\;\;\;\;+\f{1}{2}\left(\f{1}{1-x}\right)_+
\left[
\left(\f{1}{1-y}\right)_+ + \left(\f{1}{1+y}\right)_+
\right]\,,\nn
\eeeq
where the plus distributions are defined as
\beeq
&&\int_0^1 dx\, G_+(x)\, f(x)
=\int_0^1 dx\,
G(x)
\left[f(x)-f(1)\right]\,,\\
&&\int_{-1}^1 dy\, f(y)
\left(
\f{1}{1\pm y}
\right)_+
=\int_{-1}^1 dy\,
\f{f(y)-f(\mp 1)}{1\pm y}\,.
\eeeq
The delta functions in the first two terms of the expansion allow us to simplify considerably the complexity of the squared matrix element, leading to a much simpler analytical phase space integration. On the other hand, the last term in Eq. (\ref{expansion}) is finite, and then the integration can be performed (numerically) in four dimensions.

We directly present the final results.
The gluon-gluon contribution to $\hat\sigma^b$ can be split in the following way
\beq\label{contrgg}
\hat\sigma^b_{gg}=
\hat\sigma^{(r)}_{gg}+\hat\sigma^{(v)}=
\hat\sigma^{(sv)}_{gg}+\hat\sigma^{(c+)}_{gg}
+\hat\sigma^{(c-)}_{gg}+\hat\sigma^{(f)}_{gg}\,,
\eeq
where the renormalized results (for $\mu_F=\mu_R=Q$) take the following form
\beeq\label{resultadosgg}
\hat\sigma^{(sv)}_{gg}&=&
\f{\hat\sigma_{\text{LO}}}{|C_{LO}|^2}\, \delta(1-x)
\bigg\{
\left(\f{\as}{2\pi}\right)\f{4}{3}\,\text{Re}(C_{LO})\nn
\\
&+&\left(\f{\as}{2\pi}\right)^2\bigg[
\text{Re}(C_{LO})\left(
\f{8\pi^2}{3}+{\cal R}^{(2)}-8(C_{H}^{(2)}-C_{HH}^{(2)})
\right)\nn\\
&+&\text{Im}(C_{LO}){\cal I}^{(2)}+{\cal V}^{(2)}
\bigg]
\bigg\}\,,
\\
\hat\sigma^{(c+)}_{gg}&=&
\hat\sigma^{(c-)}_{gg}=
\f{\hat\sigma_{\text{LO}}}{|C_{LO}|^2}\,
\left(\f{\as}{2\pi}
\right)^2 
8\left[1-(1-x)x\right]^2 \nn\\
&\times&
\left[
2\left(
\f{\log(1-x)}{1-x}
\right)_+
-\f{\log x}{1-x}
\right]
\text{Re}(C_{LO})\,,\nn
\\
\hat\sigma^{(f)}_{gg}&=&
\int d\cos\theta_1\,d\theta_2\,dy\,
\f{\sqrt{x(x-4M_H^2/s)}}
{1024\,\pi^4}
\left(\f{1}{1-x}
\right)_+\nn\\
&\times&\left[
\left(\f{1}{1-y}\right)_+
+\left(\f{1}{1+y}\right)_+
\right]
f_{gg}(x,y,\theta_1,\theta_2)\nn\,.
\eeeq
We have already included, in these expressions, the counter terms arising from collinear factorization.
The expressions for ${\cal R}^{(2)}$, ${\cal I}^{(2)}$ and ${\cal V}^{(2)}$ can be found in Ref. \cite{deFlorian:2013uza}. We subtracted the term in ${\cal R}^{(2)}$ proportional to $C_{H}^{(2)}-C_{HH}^{(2)}$ since it has already been included in $\hat\sigma^a_{gg}$.
The expression for $f_{gg}(x,y,\theta_1,\theta_2)$ can be found in the appendix.
We also included the NLO contribution to $\hat\sigma^b$ in the definition of $\hat\sigma_{gg}^{(sv)}$.

Using a similar procedure we obtain the results for the $qg$ and $gq$ channels ($q$ stands for any massless quark or antiquark), which can be split into two contributions,
\beeq
\hat\sigma^b_{qg}&=&
\hat\sigma^{(r)}_{qg}=
\hat\sigma^{(c+)}_{qg}+\hat\sigma^{(f)}_{qg}\,,\\
\hat\sigma^b_{gq}&=&
\hat\sigma^{(r)}_{gq}=
\hat\sigma^{(c-)}_{gq}+\hat\sigma^{(f)}_{gq}\,,\nn
\eeeq
which take the following form
\beeq
\label{resultadosqg}
\hat\sigma^{(c+)}_{qg}&=&
\hat\sigma^{(c-)}_{gq}=
\f{\hat\sigma_{\text{LO}}}{|C_{LO}|^2}\,
\left(\f{\as}{2\pi}
\right)^2 
\f{16}{9}\,\Big\{\left[1+(1-x)^2\right]\nn\\
&\times&
\left[
2\log(1-x)-\log x
\right]
+x^2\Big\}
\,\text{Re}(C_{LO})\,,\nn\\
\hat\sigma^{(f)}_{qg}&=&
\int d\cos\theta_1\,d\theta_2\,dy\,
\f{\sqrt{x(x-4M_H^2/s)}}
{512\,\pi^4}\nn\\
&\times&
\left(\f{1}{1-y}\right)_+
f_{qg}(x,y,\theta_1,\theta_2)\,,\\
\hat\sigma^{(f)}_{gq}&=&
\int d\cos\theta_1\,d\theta_2\,dy\,
\f{\sqrt{x(x-4M_H^2/s)}}
{512\,\pi^4}\nn\\
&\times&
\left(\f{1}{1+y}\right)_+
f_{gq}(x,y,\theta_1,\theta_2)\nn\,.
\eeeq
Again, we already included the counter terms in the definition of $\hat\sigma^{(c+)}_{qg}$ and $\hat\sigma^{(c-)}_{gq}$.
Finally, for the quark-antiquark subprocess we have
\beq
\hat\sigma^{b}_{q\bar q}=
\int d\cos\theta_1\,d\theta_2\,dy\,
\f{\sqrt{x(x-4M_H^2/s)}}
{512\,\pi^4}\,
f_{q\bar q}(x,y,\theta_1,\theta_2)\,.
\label{resultadosqqbar}
\eeq
The expressions for $f_{qg}$, $f_{gq}$ and $f_{q\bar q}$ can be found in the appendix.

Summarizing, Eqs. (\ref{sigma_a}), (\ref{resultadosgg}), (\ref{resultadosqg}) and (\ref{resultadosqqbar}) contain all the contributions to the partonic cross section up to NNLO accuracy. We find agreement with Ref. \cite{Dawson:1998py} with respect to the NLO results.\footnote{We notice that the exact LO is taken into account in a slightly different way in Ref. \cite{Dawson:1998py}. The numerical effect is anyway small.}

\section{Phenomenology}

We present, here, the phenomenological results for the LHC.
In all cases we use the MSTW2008 \cite{Martin:2009iq} sets of parton distributions and QCD coupling at each corresponding order. The bands are obtained by varying independently the factorization and renormalization scales in the range $0.5\,Q\leq \mu_F,\mu_R \leq 2\,Q$, with the constraint $0.5\leq \mu_F/\mu_R \leq 2$.
We recall that we always normalize our results with the exact top- and bottom-mass dependence at LO.
We use $M_H=126\,\text{GeV}$, $M_t=173.18\,\text{GeV}$ and $M_b=4.75\,\text{GeV}$.

Given that at one-loop order the corrections to the effective vertex $ggHH$ are the same than those of $ggH$, we will assume for the phenomenological results that $C_{HH}^{(2)}=C_{H}^{(2)}$. We analyzed the impact of this still unknown coefficient varying its value in the range $0\leq C_{HH}^{(2)}\leq 2C_{H}^{(2)}$ and found a variation in the total cross section of less than $2.5\%$.

\begin{figure}
\begin{center}
\begin{tabular}{c}
\epsfxsize=8.7truecm
\epsffile{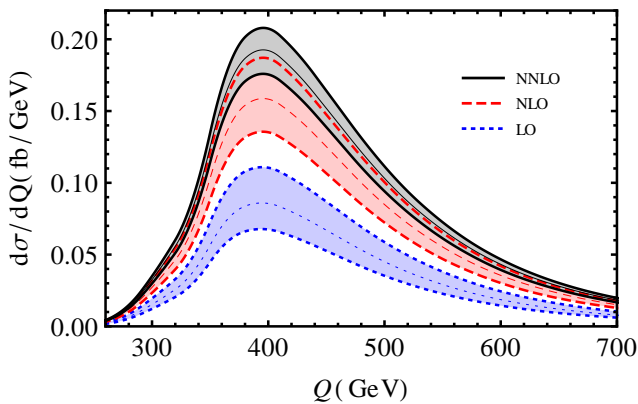}\\
\end{tabular}
\end{center}
\vspace{-0.7cm}
\caption{\label{q2distr}
Higgs pair invariant mass distribution at LO (dotted blue), NLO (dashed red) and NNLO (solid black) for the LHC at c.m. energy $E_{cm}=14\,\text{TeV}$. The bands are obtained by varying $\mu_F$ and $\mu_R$ in the range $0.5\,Q\leq \mu_F,\mu_R \leq 2\,Q$ with the constraint $0.5\leq \mu_F/\mu_R \leq 2$.}
\end{figure}

In Figure \ref{q2distr} we show the hadronic cross section for the LHC as a function of the Higgs pair invariant mass, for a c.m. energy $E_{cm}=\sqrt{s_H}=14\,\text{TeV}$, at LO, NLO and NNLO accuracy. We can observe that it is only at this order that the first sign of convergence of the perturbative series appears, finding a nonzero overlap between the NLO and NNLO bands.
Second order corrections are sizeable, this is noticeable already at the level of the total inclusive cross sections
\beeq
\sigma_{\text{LO}}&=& 17.8^{\,+5.3}_{\,-3.8}\,\text{fb}\nn\\
\sigma_{\text{NLO}}&=& 33.2^{\,+5.9}_{\,-4.9}\,\text{fb}\\
\sigma_{\text{NNLO}}&=& 40.2^{\,+3.2}_{\,-3.5}\,\text{fb}\nn
\eeeq
where the uncertainty arises from the scale variation. The increase with respect to the NLO result is then of ${\cal O}(20\%)$, and the $K$ factor with respect to the LO prediction is about $K_{\text{NNLO}}=2.3$.
The scale dependence is clearly reduced at this order, resulting in a variation of about $\pm8\%$ around the central value, compared to a total variation of ${\cal O}(\pm20\%)$ at NLO.

\begin{figure}
\begin{center}
\begin{tabular}{c}
\epsfxsize=8.7truecm
\epsffile{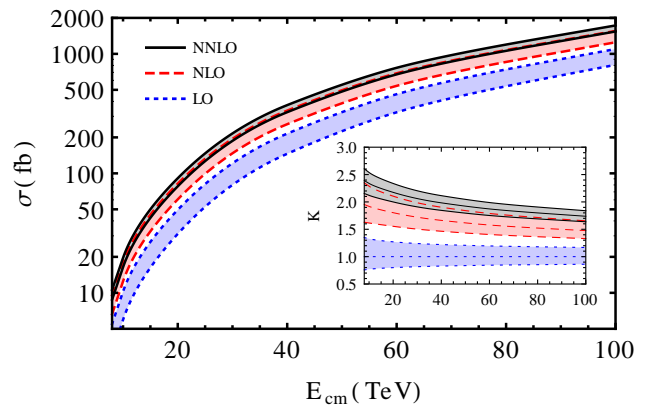}\\
\end{tabular}
\end{center}
\vspace{-0.7cm}
\caption{\label{sh}
Total cross section as a function of the c.m. energy $E_{cm}$ for the LO (dotted blue), NLO (dashed red) and NNLO (solid black) prediction. The bands are obtained by varying $\mu_F$ and $\mu_R$ as indicated in the main text. The inset plot shows the corresponding $K$ factors.}
\end{figure}

In Figure \ref{sh} we present the total cross section as a function of the c.m. energy $E_{cm}$, in the range from $8\,\text{TeV}$ to $100\,\text{TeV}$. We can observe that the size of the perturbative corrections is smaller as the c.m. energy increases. Again, in the whole range of energies the scale dependence is substantially reduced when we consider the second order corrections.

In Table \ref{tabla} we show the value of the NNLO cross section for $E_{cm}=8$, 14, 33 and $100\,\text{TeV}$. We considered three different sources of theoretical uncertainties: missing higher orders in the QCD perturbative expansion, which are estimated by the scale variation as indicated before, and uncertainties in the determination of the parton distributions and strong coupling.
To estimate the parton flux and coupling constant uncertainties we used the MSTW2008 $90\%$ C.L. error PDF sets \cite{Martin:2009bu}, which are known to provide very close results to the PDF4LHC working group recommendation for the envelope prescription \cite{Botje:2011sn}. We observe that nonperturbative and perturbative uncertainties are of the same order.

\begin{table}
\begin{center}
\begin{tabular}{l  c  c  c  c }
\hline\hline
$E_{cm}$ & $8\text{ TeV}$ & $14\text{ TeV}$ & $33\text{ TeV}$ & $100\text{ TeV}$ \\
\hline
$\sigma_{\text{NNLO}}$ & $9.76\text{ fb}$ & $40.2\text{ fb}$ & $243\text{ fb}$ & $1638\text{ fb}$ \\
Scale $[\%]$ &\footnotesize $+9.0-9.8\,$ &\footnotesize $\,+8.0-8.7\,$ &\footnotesize $\,+7.0-7.4\,$ &\footnotesize $\,+5.9-5.8$ \\ 
PDF $[\%]$ &\footnotesize $+6.0-6.1\,$ &\footnotesize $\,+4.0-4.0\,$ &\footnotesize $\,+2.5-2.6\,$ &\footnotesize $\,+2.3-2.6$ \\ 
PDF$+\as$  $[\%]$ &\footnotesize $+9.3-8.8\,$ &\footnotesize $\,+7.2-7.1\,$ &\footnotesize $\,+6.0-6.0\,$ &\footnotesize $\,+5.8-6.0$ \\ 
\hline\hline
\end{tabular}
\caption{Total cross section as a function of the c.m. energy at NNLO accuracy. We use the exact LO prediction to normalize our results. The different sources of theoretical uncertainties are discussed in the main text.\label{tabla}}
\end{center}
\end{table}

The ratio between NNLO and NLO predictions as a function of the c.m. energy is quite flat. In order to ease the use of our NNLO results, we provide the following approximated analytic expression for the $K$ factor, valid in the range $8\,\text{TeV}\leq E_{cm}\leq 100\,\text{TeV}$:
\beq
\f{\sigma_{\text{NNLO}}}{\sigma_{\text{NLO}}}=
1.149
-0.326\left(\f{E_{cm}}{1\,\text{TeV}}\right)^{-1}
+0.327\left(\f{E_{cm}}{1\,\text{TeV}}\right)^{-1/2}\,,
\eeq
which runs from $1.22$ at $8\,\text{TeV}$ to $1.18$ at $100\,\text{TeV}$.
On the other hand, the ratio between NNLO and LO runs from $2.39$ to $1.74$ in the same range of energies, and can be parametrized by the following expression
\beq
\f{\sigma_{\text{NNLO}}}{\sigma_{\text{LO}}}=
1.242
-7.17\left(\f{E_{cm}}{1\,\text{TeV}}\right)^{-1}
+5.77\left(\f{E_{cm}}{1\,\text{TeV}}\right)^{-1/2}\,.
\eeq
Finally, the total scale variation at NNLO is approximately given by $\pm p(E_{cm})\%$, with
\beq
p(E_{cm})=
4.07
-9.8\left(\f{E_{cm}}{1\,\text{TeV}}\right)^{-1}
+18.6\left(\f{E_{cm}}{1\,\text{TeV}}\right)^{-1/2}\,.
\eeq
In this case, we have $\pm 9.4\%$ and $\pm 5.8\%$ at $8$ and $100\,\text{TeV}$ respectively.

It is worth noticing that the soft-virtual approximation presented in \cite{deFlorian:2013uza} gives an extremely accurate prediction for the NNLO cross section, overestimating for example the $E_{cm}=14\,\text{TeV}$ result by less than $2\%$. As expected, this approximation works even better than for single Higgs production, due to the larger invariant mass of the final state.

\section*{Acknowledgements}

This work was supported in part by UBACYT, CONICET, ANPCyT and the Research Executive Agency (REA) of the European Union under the Grant Agreement number PITN-GA-2010-264564 (LHCPhenoNet).


\onecolumngrid
\newpage

\appendix
\section{One-loop real corrections}
\renewcommand{\theequation}{{\rm{A}}.\arabic{equation}}
\setcounter{equation}{0}

We present here the contribution to the squared matrix element arising from the class of diagrams shown in Figure \ref{diagramas}, that is, single real emission diagrams containing three effective vertices $ggH$.

Let $p_1$ and $p_2$ be the momenta of the incoming partons, $k_1$ and $k_2$ the momenta of the Higgs bosons and $k$ the momentum of the outgoing parton. We will express the matrix elements in terms of the following invariants \cite{Frixione:1993yp}
\beeq
&&\;s = (p_1+p_2)^2\;,\;\;\;\;\;\;\;\;\;
s_2 = (k_1+k_2)^2=s+t_k+u_k\;,\nn\\
&&t_k = (p_1-k)^2\;,\;\;\;\;\;\;\;\;\;
\;\hat{q}_1 = (p_1-k_2)^2=2M_H^2-s-t_k-q_1\;,\nn\\
&&\!u_k = (p_2-k)^2\;,\;\;\;\;\;\;\;\;\;
\;\hat{q}_2 = (p_2-k_1)^2=2M_H^2-s-u_k-q_2\;,\\
&&q_1 = (p_1-k_1)^2\;,\;\;\;\;\;\;\;\;\;
\!w_1 = (k+k_1)^2=M_H^2-q_1+q_2-t_k\;,\nn\\
&&q_2 = (p_2-k_2)^2\;,\;\;\;\;\;\;\;\;\;
\!w_2 = (k+k_2)^2=M_H^2+q_1-q_2-u_k\;.\nn
\eeeq
These can be written in terms of the variables $x$, $y$, $\theta_1$ and $\theta_2$ using the following expressions
\beeq
t_k&=&-\tfrac{1}{2}s(1-x)(1-y)\,,
\\
u_k&=&-\tfrac{1}{2}s(1-x)(1+y)\,,\nn\\
q_1&=&
M_H^2-\tfrac{1}{2}(s+t_k)(1-\beta_x \cos\theta_1)\,,\nn\\
q_2&=&
M_H^2-\tfrac{1}{2}(s+u_k)(1+\beta_x \cos\theta_2\sin\theta_1\sin\psi
+\beta_x\cos\theta_1\cos\psi)\,,\nn
\eeeq
where we defined the following quantities
\beeq
\beta_x&=&\sqrt{1-\f{4M_H^2}{x\,s}}\,,\\
\cos\psi&=&1-\frac{8 x}{(1+x)^2-(1-x)^2 y^2}\,.\nn
\eeeq

We start with the partonic subprocess $gg\to HHg$. The contribution to the squared matrix element is 
\beq
{\cal M}_{gg}=\f{1}{2s}\left(\f{1}{2}\right)^2\left(\f{1}{8}\right)^2\f{1}{2}
\sum_{\text{spin, color}}
\left[
{\cal A}_{gg}^{2V}\left({\cal A}_{gg}^{1V}\right)^*+
{\cal A}_{gg}^{1V}\left({\cal A}_{gg}^{2V}\right)^*
\right]\;,
\eeq
where we include the flux factor, the average over spins and colors and the factor $\tfrac{1}{2}$ for identical particles in the final state. The amplitude ${\cal A}_{gg}^{2V}$ originates from diagrams with two vertices $ggH$ (top left of Figure \ref{diagramas}), while ${\cal A}_{gg}^{1V}$ represents those diagrams with one effective vertex (top right of Figure \ref{diagramas}). This contribution can be cast in the following way
\beeq
{\cal M}_{gg}&=&\frac{\as^4 G_F^2 \text{Re}(C_{LO})}{576 \pi ^2 s}
\big[F(s,q_1,q_2,t_k,u_k)+F(s,\hat{q}_1,\hat{q}_2,t_k,u_k)\\
&+&F(s,q_2,q_1,u_k,t_k)+F(s,\hat{q}_2,\hat{q}_1,u_k,t_k)
+F(t_k,q_1,w_2,s,u_k)+F(t_k,\hat{q}_1,w_1,s,u_k)\big]\;,\nn
\eeeq
where we have defined
\beeq
F(s,q_1,q_2,t_k,u_k)&=&f_1(s,q_1,q_2,t_k,u_k)
+f_2(s,q_1,q_2,t_k,u_k)\;.
\eeeq
The function $f_1$ can be written in the following way
\beeq
&&f_1(s,q_1,q_2,t_k,u_k)=
\f{1}{q_1 s\, t_k (M_H^2 + q_1 - q_2 - u_k) u_k}
\Big[
s t_k (-q_2^2 (2 s-3 t_k) (s+t_k)+q_1 q_2 (6 s^2+3 s t_k+2 t_k^2+q_2 (s+t_k))\nn\\&&\;\;-q_1^2
(q_2 s+4 (s^2+s t_k+t_k^2)))+(2 (q_1-q_2) s^2 (2 q_1^2-2 q_1 q_2+q_2^2+s^2)+s
(-q_1^2 s+q_2 (-3 q_2^2+2 q_2 s-8 s^2)\nn\\&&\;\;+q_1 (6 q_2^2+3 q_2 s+14 s^2)) t_k+(-8
q_1^2 s-q_2 (q_2^2-3 q_2 s+7 s^2)+q_1 (q_2^2+10 q_2 s+17 s^2)) t_k^2+(-4
q_1^2+6 q_1 (q_2+s)\nn\\&&\;\;+q_2 (q_2+4 s)) t_k^3) u_k+(2 s (2 q_1^3-2 q_1^2 (q_2+s)-2
q_2 s (q_2+s)+q_1 (q_2+s) (q_2+3 s))+(q_1 (q_1-q_2) q_2+(11 q_1-6 q_2)
q_2 s\nn\\&&\;\;+2 (11 q_1-3 q_2) s^2-2 s^3) t_k+(-4 q_1^2+7 q_1 q_2-3 q_2^2+23 q_1 s+q_2
s-6 s^2) t_k^2+(6 q_1+2 q_2+s) t_k^3) u_k^2\nn\\&&\;\;+(-4 s (q_1^2+q_2 s-q_1 (q_2+2
s))-(3 q_1^2-13 q_1 s+s (7 q_2+4 s)) t_k+(6 q_1-3 q_2-2 s) t_k^2+t_k^3) u_k^3\\&&\;\;+(q_1-t_k)
(4 s+t_k) u_k^4-M_H^6 s (t_k+u_k) (t_k+2 u_k)+M_H^4 (s t_k ((-q_1+q_2)
s+(q_1+2 q_2-2 s) t_k+3 t_k^2)\nn\\&&\;\;+(2 (q_1-q_2) s^2+3 (2 q_1+q_2) s t_k+(q_1-q_2+9
s) t_k^2+5 t_k^3) u_k+(q_1 (6 s+t_k)+t_k (-q_2+9 s+6 t_k)) u_k^2+t_k u_k^3)\nn\\&&\;\;+M_H^2
(s t_k ((q_1-q_2) (q_1+q_2-2 s) s+(-q_2 (2 q_1+q_2)+(q_1+q_2) s) t_k+2
(q_1-3 q_2) t_k^2)\nn\\&&\;\;-(4 q_1 (q_1-q_2) s^2+s (-3 q_2 (-4 q_1+q_2)+(q_1+3
q_2) s-2 s^2) t_k+2 (-q_2^2+6 q_2 s-4 s^2\nn\\&&\;\;+q_1 (q_2+s)) t_k^2+2 (q_1+3 q_2+2
s) t_k^3) u_k-(2 s (4 q_1^2-2 q_1 q_2+q_2^2+s^2)+(q_1^2-q_2 (q_2-3
s)+11 q_1 s) t_k\nn\\&&\;\;+(3 (q_1+q_2)+8 s) t_k^2+6 t_k^3) u_k^2+(-4 s (q_2+s)-6 s t_k-5
t_k^2+2 q_1 (2 s+t_k)) u_k^3-(4 s+t_k) u_k^4)
\Big]\nn\;,
\eeeq
while $f_2$ is defined by
\beeq
&&f_2(s,q_1,q_2,t_k,u_k)=
\f{1}{q_2 s\, t_k^2 u_k}
\Big[
s t_k (4 q_2 s^3-s ((M_H^2+3 q_1-4 q_2) (q_1-q_2)+(4 M_H^2+q_1-11 q_2) s) t_k\nn\\&&\;\;-((M_H^2+3
q_1-4 q_2) (M_H^2-q_2)+(7 M_H^2+2 q_1-11 q_2) s) t_k^2+4 (-M_H^2+q_2) t_k^3)-(4
(q_1-q_2)^2 s^3\nn\\&&\;\;+s^2 (4 (M_H^2-q_2) (q_1-q_2)+5 (q_1-3 q_2) s) t_k+s (5 M_H^4+6 q_1^2+5
q_2 (q_2-5 s)+q_1 (-6 q_2+s)\\&&\;\;+M_H^2 (-6 q_1-4 q_2+4 s)) t_k^2+((M_H^2-q_2) (4 M_H^2-3
q_1-q_2)+3 (3 M_H^2+2 q_1-5 q_2) s+s^2) t_k^3\nn\\&&\;\;+(M_H^2-q_2+4 s) t_k^4) u_k-(-8
(M_H^2-q_1) (q_1-q_2) s^2+s (4 (M_H^2-q_1) (M_H^2-q_2)+(-5 M_H^2+8 q_1-15 q_2) s) t_k\nn\\&&\;\;+((M_H^2-q_1)
(4 M_H^2-3 q_1-q_2)+(M_H^2-q_1-20 q_2) s+5 s^2) t_k^2+2 (M_H^2+2 q_1-4 q_2) t_k^3+t_k^4)
u_k^2\nn\\&&\;\;+(-4 (M_H^2-q_1)^2 s+(3 M_H^2-3 q_1+10 q_2) s t_k+(-5 M_H^2+q_1+10 q_2+s) t_k^2+t_k^3)
u_k^3+4 q_2 t_k u_k^4\nn
\Big]\;.
\eeeq
Finally, the function $f_{gg}(x,y,\theta_1,\theta_2)$ in Eq. (\ref{resultadosgg}) is defined in terms of ${\cal M}_{gg}$ as
\beq
f_{gg}(x,y,\theta_1,\theta_2) =
\Delta_{LO}\,
s (1-x)^2 (1-y^2){\cal M}_{gg}\,,
\eeq
where the factor $\Delta_{LO}$ introduces the exact LO normalization into our results (which are calculated in the large top-mass limit) and it is given by
\beq
\Delta_{LO} =
\f{
\left| C_\triangle F_\triangle + C_\square F_\square \right|^2
+ \left| C_\square G_\square \right|^2
}{|\f{2}{3}C_{LO}|^2}\,.
\eeq
In the large $M_t$ limit we have $\Delta_{LO}\to 1$.
This factor depends on $Q^2=x\,s$ and the Mandelstam invariant $t$, which is given by
\beq
t=-\f{1}{2}\left(Q^2-2M_H^2- Q\sqrt{Q^2-4M_H^2}\cos\theta_1\right)\,.
\eeq

For the partonic subprocess $qg\to HHq$ we define the contribution to the squared matrix element in a similar way we did for ${\cal M}_{gg}$
\beq
{\cal M}_{qg}=\f{1}{2s}\left(\f{1}{2}\right)^2\f{1}{3}\,\f{1}{8}\,\f{1}{2}
\sum_{\text{spin, color}}
\left[
{\cal A}_{qg}^{2V}\left({\cal A}_{qg}^{1V}\right)^*+
{\cal A}_{qg}^{1V}\left({\cal A}_{qg}^{2V}\right)^*
\right]\;.
\eeq
This contribution can be written as
\beq
{\cal M}_{qg}=\frac{\as^4 G_F^2 \text{Re}(C_{LO})}{648 \pi ^2 s}
\left[
h(s,q_1,q_2,t_k,u_k)+h(s,\hat{q}_1,\hat{q}_2,t_k,u_k)
\right]\;,
\eeq
where the function $h$ is defined by the expression
\beeq
&&h(s,q_1,q_2,t_k,u_k)=\f{1}{t_k^2 q_2}\Big[
2 (-M_H^4 t_k^2-q_2^2 (s^2+s t_k+t_k^2)+M_H^2 t_k (-M_H^2+t_k) u_k-(M_H^2-t_k)^2
u_k^2\nn\\&&\;\;-q_1^2 (s+u_k)^2+q_1 (s (-M_H^2 t_k+q_2 (2 s+t_k))+(2 (M_H^2+q_2) s+(M_H^2-q_2-2
s) t_k) u_k\\&&\;\;+2 (M_H^2-t_k) u_k^2)+q_2 (-t_k (s^2+u_k (t_k+u_k))+M_H^2
(s (t_k-2 u_k)+t_k (2 t_k+u_k))))
\Big]\;.\nn
\eeeq
For ${\cal M}_{gq}$ and ${\cal M}_{q\bar q}$ we obtain via crossings the following results
\beeq
{\cal M}_{gq}&=&\frac{\as^4 G_F^2 \text{Re}(C_{LO})}{648 \pi ^2 s}
\left[
h(s,\hat q_2,\hat q_1,u_k,t_k)+h(s,q_2,q_1,u_k,t_k)
\right]\;,\\
{\cal M}_{q\bar q}&=&-\frac{\as^4 G_F^2 \text{Re}(C_{LO})}{243 \pi ^2 s}
\left[
\, h(t_k,q_1,w_2,s,u_k)+\, h(t_k,\hat q_1,w_1,s,u_k)
\right]\;,
\eeeq
Finally, the functions $f_{qg}$, $f_{gq}$ and $f_{q\bar q}$ are defined as
\beeq
f_{qg}(x,y,\theta_1,\theta_2) &=&
\Delta_{LO}\,
s (1-x) (1-y){\cal M}_{qg}\,,\nn\\
f_{gq}(x,y,\theta_1,\theta_2) &=&
\Delta_{LO}\,
s (1-x) (1+y){\cal M}_{gq}\,,\\
f_{q\bar q}(x,y,\theta_1,\theta_2) &=&
\Delta_{LO}\,
s (1-x) {\cal M}_{q\bar q}\,.\nn
\eeeq
We recall that $q$ stands for any massless quark or antiquark flavour.


\begin{thebibliography}{90}


\bibitem{Aad:2012tfa}
  G.~Aad {\it et al.}  [ATLAS Collaboration],
  Phys.\ Lett.\ B {\bf 716} (2012) 1
  [arXiv:1207.7214 [hep-ex]].


\bibitem{Chatrchyan:2012ufa}
  S.~Chatrchyan {\it et al.}  [CMS Collaboration],
  Phys.\ Lett.\ B {\bf 716} (2012) 30
  [arXiv:1207.7235 [hep-ex]].


\bibitem{Englert:1964et}
  F.~Englert and R.~Brout,
  Phys.\ Rev.\ Lett.\  {\bf 13} (1964) 321;
  P.~W.~Higgs,
  Phys.\ Lett.\  {\bf 12} (1964) 132,
  Phys.\ Rev.\ Lett.\  {\bf 13} (1964) 508.







\bibitem{Baur:2002qd}
  U.~Baur, T.~Plehn and D.~L.~Rainwater,
  Phys.\ Rev.\ D {\bf 67} (2003) 033003
  [hep-ph/0211224].
  
\bibitem{Dolan:2012rv}
  M.~J.~Dolan, C.~Englert and M.~Spannowsky,
  JHEP {\bf 1210} (2012) 112
  [arXiv:1206.5001 [hep-ph]].
  
\bibitem{Papaefstathiou:2012qe}
  A.~Papaefstathiou, L.~L.~Yang and J.~Zurita,
  Phys.\ Rev.\ D {\bf 87} (2013) 011301
  [arXiv:1209.1489 [hep-ph]].

  
\bibitem{Baglio:2012np}
  J.~Baglio, A.~Djouadi, R.~Grober, M.~M.~Muhlleitner, J.~Quevillon and M.~Spira,
  arXiv:1212.5581 [hep-ph].
  
\bibitem{Baur:2003gp}
  U.~Baur, T.~Plehn and D.~L.~Rainwater,
  Phys.\ Rev.\ D {\bf 69} (2004) 053004
  [hep-ph/0310056].
  
  
\bibitem{Dolan:2012ac}
  M.~J.~Dolan, C.~Englert and M.~Spannowsky,
  Phys.\ Rev.\ D {\bf 87} (2013) 055002
  [arXiv:1210.8166 [hep-ph]].

\bibitem{Goertz:2013kp}
  F.~Goertz, A.~Papaefstathiou, L.~L.~Yang and J.~Zurita,
  arXiv:1301.3492 [hep-ph].
  
\bibitem{Shao:2013bz}
  D.~Y.~Shao, C.~S.~Li, H.~T.~Li and J.~Wang,
  arXiv:1301.1245 [hep-ph].


  
\bibitem{Gouzevitch:2013qca} 
  M.~Gouzevitch, A.~Oliveira, J.~Rojo, R.~Rosenfeld, G.~Salam and V.~Sanz,
  arXiv:1303.6636 [hep-ph].



\bibitem{Glover:1987nx}
  E.~W.~N.~Glover and J.~J.~van der Bij,
  Nucl.\ Phys.\ B {\bf 309} (1988) 282.


\bibitem{Eboli:1987dy}
  O.~J.~P.~Eboli, G.~C.~Marques, S.~F.~Novaes and A.~A.~Natale,
  Phys.\ Lett.\ B {\bf 197} (1987) 269.




\bibitem{Plehn:1996wb}
  T.~Plehn, M.~Spira and P.~M.~Zerwas,
  Nucl.\ Phys.\ B {\bf 479} (1996) 46
   [Erratum-ibid.\ B {\bf 531} (1998) 655]
  [hep-ph/9603205].

\bibitem{Dawson:1998py}
  S.~Dawson, S.~Dittmaier and M.~Spira,
  Phys.\ Rev.\ D {\bf 58} (1998) 115012
  [hep-ph/9805244].


  
 
\bibitem{deFlorian:2013uza}
  D.~de Florian and J.~Mazzitelli,
  Phys.\ Lett.\ B {\bf 724} (2013) 306
  [arXiv:1305.5206 [hep-ph]].
  
  
\bibitem{deFlorian:2012za}
  D.~de Florian and J.~Mazzitelli,
  JHEP {\bf 1212} (2012) 088
  [arXiv:1209.0673 [hep-ph]].
  
  
\bibitem{Grigo:2013rya}
  J.~Grigo, J.~Hoff, K.~Melnikov and M.~Steinhauser,
  arXiv:1305.7340 [hep-ph].
  
  
\bibitem{Kramer:1996iq}
  M.~Kramer, E.~Laenen and M.~Spira,
  Nucl.\ Phys.\ B {\bf 511} (1998) 523.

\bibitem{Chetyrkin:1997iv} 
  K.~G.~Chetyrkin, B.~A.~Kniehl and M.~Steinhauser,
  Phys.\ Rev.\ Lett.\  {\bf 79}, 353 (1997)
  [hep-ph/9705240].
  
  
\bibitem{Djouadi:1991tk}
A.~Djouadi, M.~Spira and P.~M.~Zerwas,
Phys.\ Lett.\ B {\bf 264} (1991) 440.
  
  
\bibitem{Harlander:2002wh}
  R.~V.~Harlander and W.~B.~Kilgore,
  Phys.\ Rev.\ Lett.\  {\bf 88} (2002) 201801.

\bibitem{Anastasiou:2002yz}
  C.~Anastasiou and K.~Melnikov,
  Nucl.\ Phys.\  B {\bf 646} (2002) 220.

\bibitem{Ravindran:2003um}
  V.~Ravindran, J.~Smith and W.~L.~van Neerven,
  Nucl.\ Phys.\  B {\bf 665} (2003) 325.
  
  
  
\bibitem{Hahn:2000kx}
  T.~Hahn,
  Comput.\ Phys.\ Commun.\  {\bf 140} (2001) 418
  [hep-ph/0012260].

\bibitem{Mertig:1990an}
  R.~Mertig, M.~Bohm and A.~Denner,
  Comput.\ Phys.\ Commun.\  {\bf 64} (1991) 345.
  
  
  
\bibitem{Frixione:1995ms}
  S.~Frixione, Z.~Kunszt and A.~Signer,
  Nucl.\ Phys.\ B {\bf 467} (1996) 399
  [hep-ph/9512328].
  
\bibitem{Frixione:1993yp}
  S.~Frixione,
  Nucl.\ Phys.\ B {\bf 410} (1993) 280.


\bibitem{Martin:2009iq}
  A.~D.~Martin, W.~J.~Stirling, R.~S.~Thorne and G.~Watt,
  Eur.\ Phys.\ J.\ C {\bf 63} (2009) 189
  [arXiv:0901.0002 [hep-ph]].

\bibitem{Martin:2009bu}
  A.~D.~Martin, W.~J.~Stirling, R.~S.~Thorne and G.~Watt,
  Eur.\ Phys.\ J.\ C {\bf 64} (2009) 653
  [arXiv:0905.3531 [hep-ph]].
  
\bibitem{Botje:2011sn}
  M.~Botje, J.~Butterworth, A.~Cooper-Sarkar, A.~de Roeck, J.~Feltesse, S.~Forte, A.~Glazov and J.~Huston {\it et al.},
  arXiv:1101.0538 [hep-ph].



\end{thebibliography}
\end{document}